\documentclass[letter]{jpsj2}

\newcommand{\Tc}{$T_{\rm c}$}
\newcommand{\invT}{$1/T_{1}$}

\begin{document}

\title{Anisotropic superconducting gap in transuranium superconductor PuRhGa$_{5}$: Ga NQR study on a single crystal}

\author{Hironori \textsc{Sakai}$^{1}$\thanks{E-mail address: piros@popsvr.tokai.jaeri.go.jp}, Yo \textsc{Tokunaga}$^{1}$, Tatsuya \textsc{Fujimoto}$^{1}$,  Shinsaku \textsc{Kambe}$^{1}$, Russell E. \textsc{Walstedt}$^{1}$, Hiroshi \textsc{Yasuoka}$^{1}$, Dai \textsc{Aoki}$^{2}$, Yoshiya \textsc{Homma}$^{2}$, Etsuji \textsc{Yamamoto}$^{1}$, Akio \textsc{Nakamura}$^{1}$, Yoshinobu \textsc{Shiokawa}$^{1,2}$, Kunihisa \textsc{Nakajima}$^{3}$, Yasuo \textsc{Arai}$^{3}$, Tatsuma D. \textsc{Matsuda}$^{1}$, Yoshinori \textsc{Haga}$^{1}$, Yoshichika \textsc{\={O}nuki}$^{1,4}$}

\inst{$^{1}$Advanced Science Research Center, Japan Atomic Energy Research Institute, Tokai, Ibaraki 319-1195, JAPAN \\
$^{2}$Institute for Materials Research, Tohoku University, Oarai, Ibaraki 311-1313, JAPAN\\
$^{3}$Department of Nuclear Energy System, Japan Atomic Energy Research Institute, Tokai, Ibaraki 319-1195, JAPAN\\
$^{4}$Graduate School of Science, Osaka University, Toyonaka, Osaka 560-0043,JAPAN}

\date{Submission date: 18 Mar. 2005: Acceptance: 30 Mar. 2005}

\abst{
$^{69,71}$Ga NMR/NQR studies
have been performed  on a single crystal
of the transuranium superconductor PuRhGa$_{5}$
with $T_{\rm c} \simeq 9$ K.
We have observed a $^{69}$Ga NQR line at $\sim$29.15 MHz,
and assigned it to the $4i$ Ga site using the NMR results.
The $^{69}$Ga NQR spin-lattice relaxation rate $1/T_{1}$ shows no coherence peak just below \Tc,
but obeys a $T^{3}$ behavior below \Tc. 
This result strongly suggests that 
PuRhGa$_{5}$ is an unconventional superconductor
having an anisotropic superconducting gap.
The gap amplitude 2$\Delta_{0}(T\rightarrow 0)\simeq5k_{\rm B}T_{\rm c}$ and the residual density of states $N_{\rm res}(T=0)/N_{\rm 0}(T=T_{\rm c})\simeq$ 0.25 have been determined assuming a simple polar function of the form $\Delta(\theta,\phi)$=$\Delta_{0}\cos\theta$, where $\theta$ and $\phi$ are angular parameters on the Fermi surface.
}
 
\kword{PuRhGa$_{5}$, Superconductivity, Ga NQR}

\maketitle

The recent discovery of Pu-based superconductors PuTGa$_{5}$ (T=Co, Rh) 
has stimulated interest in further experiments on transuranium compounds.
A lot of attention is now focused on these Pu heavy-fermion compounds,
which have relatively high superconducting (SC) transition temperatures,
{\it i.e.,} \Tc$\simeq$ 18 K \cite{sarrao} (PuCoGa$_{5}$)
and \Tc$\simeq$ 9 K \cite{wastin} (PuRhGa$_{5}$).
The PuTGa$_{5}$ systems belong to a large family
of ``115 compounds", which all crystallize
in the same quasi-two-dimensional structure.
In this family,
the Ce115 isomorphs CeT'In$_{5}$ (T'=Co, Ir, Rh)
\cite{cecoin5,ceirin5,cerhin5},
which are well-known 4{\it f} heavy fermion systems with antiferromagnetic tendencies,
show $d$-wave superconductivity
with relatively low \Tc\ values: \Tc=0.4 K and 2.3 K
for CeIrIn$_{5}$ and CeCoIn$_{5}$, respectively.\cite{cecoin5kohori,ceirin5zheng,cerh115mito,ce115curro}
On the other hand,
the U115 \cite{tokiwa,ikeda,tokiwa2} and Np115 \cite{colineau,aoki,aoki2,aoki3} compounds usually show itinerant
magnetic order or Pauli paramagnetism, but up to now have not shown any superconductivity.
Band calculations \cite{maehira,maehira2}
and de Haas-van Alphen experiments \cite{tokiwa,tokiwa3,ikeda2,aoki,aoki2}
have revealed that An115 (An=U, Np, Pu,$\cdots$) compounds often have
cylindrical Fermi surfaces, reflecting a quasi-two-dimensional character.

To elucidate the relatively high \Tc\ values exhibited by PuTGa$_5$,
it is important to determine the pairing symmetry,
which reflects the SC pairing mechanism. 
Nuclear Magnetic Resonance (NMR and NQR) is a
microscopic probe well-suited to this task.
In particular,
the $T$ dependence of
NQR (zero-field) spin-lattice relaxation rates (\invT) can
provide crucial information regarding the SC pairing symmetry.
In this letter,
we report 
$^{69,71}$Ga NMR/NQR experimental results
above and below the SC transition in PuRhGa$_{5}$.

A single crystal of PuRhGa$_{5}$ has been prepared using the Ga flux method. \cite{haga}
The dimensions of the single crystal are $1\times 2\times 3$ mm$^3$.
This crystal has been attached with varnish to a silver cap for thermal contact, coated with epoxy resin,
and sealed tightly in a polyimid tube in order to
avoid radiation contamination.
The sealed sample was then mounted into an rf coil.
NQR/NMR measurements have been carried out
in the temperature range 1.5-300 K
using a phase-coherent, pulsed spectrometer,
which has been installed in the radiation controlled area. 
$T_{1}$ was measured
for the Ga NQR lines using the saturation-recovery method.
The recovery
of nuclear magnetization $M(t)$ from a saturation pulse comb
was single-exponential type,
$\{M(t)-M(\infty)\}/M(\infty) \propto \exp(-3t/T_{1})$,
in the whole temperature range.
The superconductivity of PuRhGa$_{5}$  has been reported  to be
influenced by self-radiation damage due to spontaneous $\alpha$ decay,
{\it i.e.},
\Tc\ decreases gradually
with $dT_{\rm c}/dt\simeq -0.2 \sim -0.5$ K/month.\cite{wastin}
Accordingly, in order to study the superconductivity of PuRhGa$_{5}$,
brief and  timely experimental measurements are required.
In our experiments,
the bulk \Tc\ was determined by both
{\it in situ} ac susceptibility ($\chi_{\rm ac}$) measurements using the rf coil,
and the sudden decrease of the NQR signal intensity, as mentioned below. 
\Tc\ was found to be 8.5 K in zero field
at the starting point of the present experiments,
which was one month after the synthesis of the single crystal.
All the NMR/NQR experiments reported here
were conducted within a single month.

\begin{figure}[tb]
\begin{center}
\includegraphics[keepaspectratio, width=5cm]{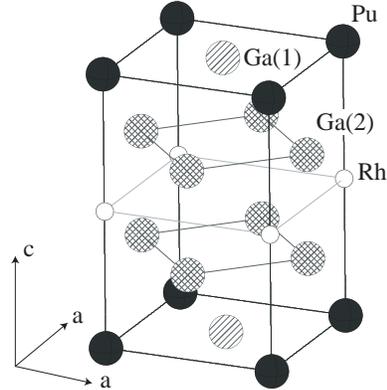}
\caption{\label{crystalstructure}Crystal structure of PuRhGa$_{5}$.}
\end{center}
\end{figure}

PuRhGa$_{5}$ crystalizes in the tetragonal HoCoGa$_{5}$ structure
with lattice parameters of $a =$ 4.2354 and $c =$ 6.7939 \AA \cite{wastin},
as shown in Fig. \ref{crystalstructure}.
This structure can be viewed as alternating PuGa$_{3}$
and RhGa$_{2}$ layers stacked along the $c$ axis.
There are two crystallographically inequivalent Ga sites,
which are denoted Ga(1) (the $1c$ site) and Ga(2) (the $4i$ site), respectively.
The Ga(1) site is surrounded by four Pu atoms in the $c$ plane.
On the other hand, the Ga(2) site is surrounded
by two Pu and two Rh atoms in the $a$ plane.
Therefore, in an applied field ($H_{0}$),
the Ga(2) sites split into two magnetically inequivalent sites,
except for the case of $H_{0}\parallel$ $c$ axis.

\begin{figure}[tb]
\begin{center}
\includegraphics[keepaspectratio, width=8cm]{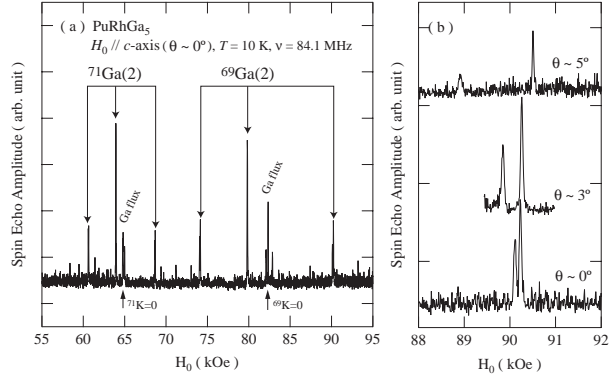}
\caption{\label{NMRspectrum}(a) Field-swept NMR spectrum in PuRhGa$_{5}$ at 10 K at a frequency of 84.1 MHz 
with the applied field ($H_{0}$) parallel to the $c$ axis. Spectral assignments are denoted by arrows, 
and the zero-shift positions $^{69,71}K=0$ are also indicated.  The two NMR lines near $^{69,71}K=0$ originate from residual Ga flux. 
(b) Enlarged spectra
around 90 kOe at a same frequency of 84.1MHz
at different angles ($\theta$)
between $H_{0}$ and the $c$ axis.
}
\end{center}
\end{figure}

Figure \ref{NMRspectrum}(a) shows
a field-swept NMR spectrum in the normal state at $T = 10$ K and a frequency of 84.1 MHz,
with $H_{0}$  parallel to the $c$-axis.
This spectrum is well explained
if we associate two sets of lines to $^{69}$Ga(2) and $^{71}$Ga(2) sites,
where each set has a center line and two quadrupolar satellite lines
denoted by arrows in Fig.\ref{NMRspectrum}.
Since the Ga(2) sites have a lower local symmetry
than the Ga(1), with four-fold axial symmetry,
each center line lies at an asymmetric position between the satellites.
This is due to the second-order quadrupolar effect.
The Knight shift ($K$),
electric field gradient parameter ($\nu_{\rm EFG}\equiv e^2qQ/2h$),
and asymmetry parameter ($\eta$) of the Ga(2) site have been estimated
from second-order perturbation theory,
where $eQ$ is nuclear quadrupole moment
and $eq$ is the largest electric field gradient. \cite{abragam}
The calculated $K$, $\nu_{\rm EFG}$ and $\eta$ at 10 K are
$K \simeq 0.19$ \%, $^{69}\nu_{\rm EFG} = 28.2$ MHz,
$^{71}\nu_{\rm EFG} = 17.4$ MHz,
and $\eta \simeq 0.42$, respectively.
Here, the ratio $^{69}\nu_{\rm EFG}/^{71}\nu_{\rm EFG} \simeq 1.6$
is consistent with that of the nuclear quadrupole moments
($^{69}Q/^{71}Q = 1.59$).
These values of $\nu_{\rm EFG}$ and $\eta$ for Ga(2) sites
are similar to those of the U115 \cite{haru} and Np115\cite{kambe,sakai} families,
{\it e.g.},
$^{69}\nu_{\rm EFG}$ and $\eta$ at the same site are 27.5 MHz and 0.2 for UPtGa$_{5}$, and $\sim$27.3 MHz and $\sim$0.29 for NpCoGa$_{5}$, respectively.

It should be noted here that
the above assignment of Ga(2) NMR lines
has been confirmed further
by the field orientation dependence ($\theta$) of the spectra
in Fig. \ref{NMRspectrum}(b),
where $\theta$ is an angle between $H_{0}$ and the $c$ axis.
As seen in Fig.\ref{NMRspectrum}(b),
a small increase of $\theta$ causes a splitting of each NMR line.
If these lines were ascribed to Ga(1) sites,
such a splitting could not be expected,
because the Ga(1) sites remain equivalent under any applied field. 
The $^{69,71}$Ga(1) NMR signals are very weak due to a short $T_{2}$,
and not seen in Fig.\ref{NMRspectrum}(a) under this experimental condition.

\begin{figure}[tb]
\begin{center}
\includegraphics[keepaspectratio, width=6cm]{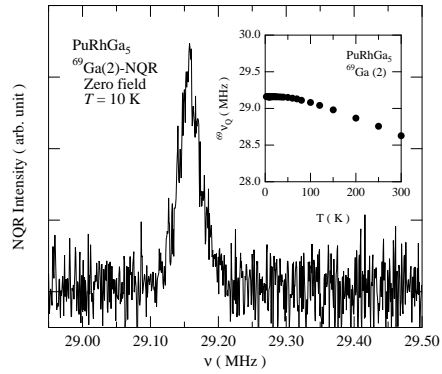}
\caption{\label{NQRspectrum}$^{69}$Ga(2) NQR spectrum with rf field parallel to $c$ axis at zero field in PuRhGa$_{2}$. 
The inset shows the temperature dependence of the $^{69}$Ga(2) NQR frequency.
}
\end{center}
\end{figure}

Next, we turn to the zero-field NQR experiments.
The $^{69}$Ga(2) NQR spectrum of the single crystal 
has been observed at $\nu_{\rm Q}\simeq 29.15$ MHz
as shown in Fig.\ref{NQRspectrum}.
Here, the value of $\nu_{\rm Q}$
is consistent with the expected frequency,
$\nu_{\rm Q} = \nu_{\rm EFG} \sqrt{1+\eta^{2}/3}$,
extracted from $\nu_{\rm EFG}$ and $\eta$ in the NMR results. 
The narrow NQR line width ($\sim$20 kHz)
assures us that this sample is still of good quality, {\it i.e.},
broadening effects due to defects and/or impurities
are not serious at this point.
The temperature dependence
of $\nu_{\rm Q}$ is shown 
in the inset of Fig.\ref{NQRspectrum}.
The value of $\nu_{\rm Q}$ increases gradually
as temperature decreases from 300 K, and saturates below $\sim$50 K.
The slight temperature dependence of $\nu_{\rm Q}$
is caused by that of $\nu_{\rm EFG}$,
because the value of $\eta$ is found
to be temperature-independent from NMR results.
Note that the $^{71}$Ga(2) NQR line has been observed
concurrently at the frequency of 18.38 MHz at $T = 10$ K.

\begin{figure}[tb]
\begin{center}
\includegraphics[keepaspectratio, width=8cm]{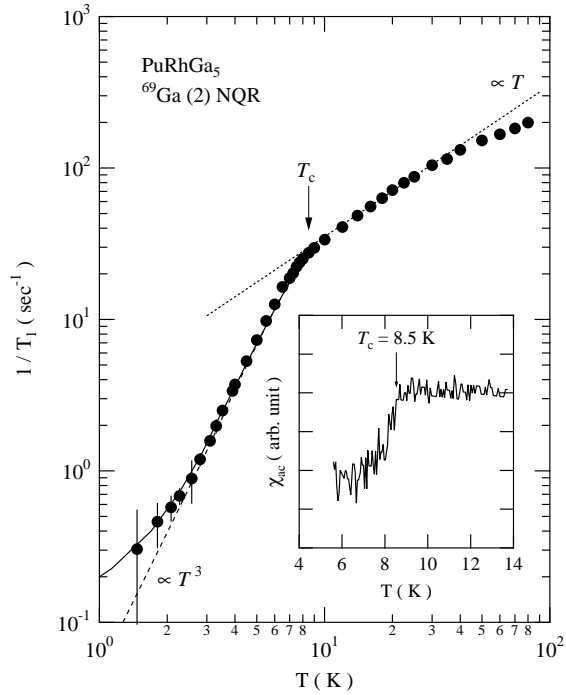}
\caption{\label{invT1}Temperature dependence of \invT\ for the $^{69}$Ga(2) NQR line in PuRhGa$_{5}$. 
The inset shows the temperature dependence
of {\it in situ} ac susceptibility using the rf coil.
The solid curve has been calculated assuming a line-node gap with $2\Delta_{0}(0)\sim 5k_{\rm B}T_{\rm c}$ and a 
residual DOS $N_{\rm res}/N_{0}\sim0.25$. The broken curve shows the same calculation without any residual DOS.
The dotted line shows Korringa-like behavior for comparison.}
\end{center}
\end{figure}

$T_{1}$ has been measured
for both $^{69}$Ga(2) and $^{71}$Ga(2) NQR lines
in order to clarify the $T_{1}$ process.
The ratio $^{69}T^{-1}/^{71}T^{-1}$ of data so obtained is found to be
nearly equal to $^{69}\gamma_{\rm n}^{2}/^{71}\gamma_{\rm n}^{2}$
below $\sim$150 K,
where $\gamma_{\rm n}$ denotes the nuclear gyromagnetic ratio.
This indicates that the magnetic dipole relaxation mechanism is dominant
over quadrupolar relaxation, at least in the low temperature range.
The inset to Fig.\ref{invT1} shows
the result of {\it in situ} $\chi_{\rm ac}$ measurements.
Here, this measurement was performed
before and after NQR $T_{1}$ measurements.
The onset \Tc\ was determined to be 8.5 K from $\chi_{\rm ac}$ data,
where the distribution of \Tc\ was estimated to be less than $\sim$0.5 K.
In the duration of a single week to obtain NQR $T_{1}$ data,
the onset \Tc\ was unchanged within less than $\sim$0.1 K.
It is noted that the NQR signal intensity was weakened suddenly
below 8.5 K, because the rf pulse was attenuated due to the Meissner effect.

Figure \ref{invT1} shows the temperature dependence of \invT\
obtained for the $^{69}$Ga(2) NQR line.
In the normal state from \Tc\ to $\sim$ 30 K,
$1/T_{1}$ is approximately proportional to $T$,
{\it i.e.},
Korringa-like behavior
associated with typical coherent Fermi liquid state.
The values of $1/T_{1}$ above $\sim$30 K
fall below Korringa-like behavior,
which is thought to be related
to a reduction in the coherence of the Fermi liquid state.
This Korringa-like behavior is also confirmed
by preliminary high-field NMR $T_{1}$ measurements at $H_{0}\sim$110 kOe ($\parallel c$),
where \Tc\ is suppressed down to $\sim$4 K. 
The present $T_{1}$ results suggest
that the SC state in PuRhGa$_{5}$ sets in
after a Fermi liquid state is established below $\sim$30 K.
Further discussion of $T_{1}$ in the normal state
will be presented elsewhere.

In the SC state,
we have found that
\invT\ shows no coherence peak just below \Tc,
but decreases $\propto$ $T^{3}$ as $T$ decreases,
as seen in Fig.\ref{invT1}.
Besides, a deviation from $T^{3}$ behavior
is also observed for $T$ well below \Tc.
The \invT\ behavior we found
cannot be explained in terms of a fully open SC gap with $s$-wave symmetry.
These results strongly suggest that PuRhGa$_{5}$
has an anisotropic gap in the SC state.
We can reproduce the behavior of \invT\ 
under the assumption of $d$-wave symmetry with line nodes,
which is widely accepted to be realized in the Ce115 family.
This calculation has been done
in the following way:
An anisotropic gap function is assumed
as a polar function $\Delta(\theta,\phi)=\Delta_{0}\cos\theta$,
where $\theta$ and $\phi$ mean angular parameters on the Fermi surface,
and the temperature dependence of $\Delta_{0}$
is assumed to be BCS-like.
Then, the temperature dependence of \invT\ below \Tc\
has been calculated from the following integral,
\[
\Bigl(\frac{1}{T_{1}}\Bigr) / \Bigl (\frac{1}{T_{1}}\Bigr)_{T=T_{\rm c}}= \frac{2}{k_{\rm B}T}\int \Bigl\langle  \frac{N_{\rm s}(E)^{2}}{N_{0}^{2}} \Bigr\rangle _{\theta, \phi}f(E)[1-f(E)] dE,
\]
where $N_{\rm s}(E)=N_{0} E/\sqrt{E^{2}-\Delta(\theta,\phi)^{2}}$ with $N_{0}$ being the density of states (DOS) in the normal state
, $f(E)$ is the Fermi distribution function,
and $\langle\cdots\rangle_{\theta,\phi}$ means the anglar average over the Fermi surface. 
From the best fit of the experimental data to this calculation,
we evaluate the SC gap
$2\Delta_{0}(T\rightarrow0) \simeq 5 k_{\rm B}$\Tc\
with a residual DOS ($N_{\rm res}/N_{\rm 0}\simeq 0.25$) in the SC state.
This gap estimate is similar to that of CeIrIn$_{5}$ with \Tc$\simeq 0.4$ K, \cite{ceirin5zheng}
while it is smaller than that of CeCoIn$_{5}$,
{\it i.e.},
$2\Delta_{0}(0)\simeq 8 k_{\rm B}T_{\rm c}$ with \Tc$\simeq 2.3$ K. 
\cite{cecoin5kohori}
In the high-\Tc\ cuprates,
the value of $2\Delta_{0}(0)$ often reaches $\sim10 k_{\rm B}T_{\rm c}$.
PuRhGa$_{5}$ would be classified
as an intermediate-coupling superconductor.

The finite residual DOS in the $d$-wave SC state is mostly caused by
potential scattering
associated with nonmagnetic impurities, distortion, and contamination with a 
secondary phase etc. \cite{ishida,kitaoka}
In Pu compounds,
the influence of unavoidable potential scattering
coming from self-radiation damage is also expected.
In the literature, the relation between the size of 
the residual DOS and the reduction rate of \Tc\  has been calculated
based on an anisotropic SC model,
where the scattering is treated in the unitarity (strong) limit.\cite{hotta,miyake} 
Using the relation with $N_{\rm res}/N_0 \simeq 0.25$ observed in our sample,
we can obtain $T_{\rm c}/T_{\rm c0} \simeq 0.94$,
where $T_{\rm c0}$ is the intrinsic value of \Tc\
without potential scattering effects.
We can estimate
the maximum (intrinsic) $T_{\rm c0} = 9.0$ K
using the $T_{\rm c}=8.5$ K in our sample observed a month after synthesis.
This $T_{\rm c0}$ is consistent with the observed \Tc\ just after synthesis.
The aging effect on the residual DOS in the SC gap
will be an interesting effect to study in Pu-based superconductors. 

In summary,
we have succeeded
in $^{69,71}$Ga NMR/NQR measurements
for the Ga(2) site of transuranium superconductor PuRhGa$_{5}$.
The $^{69,71}$Ga(2) NQR lines have been found at
frequencies consistent with an analysis based on NMR results.
The NQR \invT\ in the SC state
shows no coherence peak just below \Tc,
but obeys a $T^{3}$ behavior below \Tc.
Such a result is strong evidence
that PuRhGa$_{5}$ is an unconventional superconductor
with an anisotropic SC gap.
Assuming a $d$-wave symmetry,
the SC gap $\Delta_{0}(0) \simeq 5k_{\rm B}T_{\rm c}$
with $N_{\rm res}/N_0 \simeq 0.25$
has been evaluated.
To determine the parity of SC pairing in PuRhGa$_{5}$,
Knight shift measurements are in progress.

We thank Dr. T. Hotta and Dr. T. Maehira for helpful discussions.
This work was supported
in part by the Grant-in-Aid for Scientific Research of MEXT (Grants. No. 14340113).

-Added note.
After completion of this paper, an NMR/NQR study on the closely related compound PuCoGa$_{5}$ by Curro {\it et al.}, has appeared \cite{curro}.
It is useful to note some contrasting points
in the behavior of these two Pu superconductors.
First, the $T_1$ results of Curro {\it et al.}, are also analyzed in terms of $d$-wave superconductivity,
finding a SC gap value 2$\Delta_0 \simeq 8k_{\rm B}T_{\rm c}$ ({\it c.f.}, $5k_{\rm B}T_{\rm c}$ for PuRhGa$_{5}$).
Secondly, the $T$-variation of $(T_1T)^{-1}$ for PuCoGa$_{5}$
shows a monotonic increase right down to $T_{\rm c}$,
whereas the Rh isomorph shows a Korringa-like behavior below $\sim$30 K.


\end{document}